\documentclass{iopart}
\usepackage{epsfig,cite}

\begin{document}

\title{Advanced localization of massive black hole coalescences with LISA}

\author{Ryan N Lang and Scott A Hughes}

\address{Department of Physics and Kavli Institute for Astrophysics
and Space Research, Massachusetts Institute of Technology, 77
Massachusetts Avenue, Cambridge, MA 02139, USA}

\ead{rlang@mit.edu}

\begin{abstract}
The coalescence of massive black holes is one of the primary sources
of gravitational waves (GWs) for LISA.  Measurements of the GWs can
localize the source on the sky to an ellipse with a major axis of a
few tens of arcminutes to a few degrees, depending on source redshift,
and a minor axis which is 2--4 times smaller.  The distance (and thus
an approximate redshift) can be determined to better than a per cent
for the closest sources we consider, although weak lensing degrades
this performance.  It will be of great interest to search this
three-dimensional `pixel' for an electromagnetic counterpart to the
GW event.  The presence of a counterpart allows unique studies which
combine electromagnetic and GW information, especially if the
counterpart is found prior to final merger of the holes.  To
understand the feasibility of early counterpart detection, we
calculate the evolution of the GW pixel with time. We find that the
greatest improvement in pixel size occurs in the final day before
merger, when spin precession effects are maximal. The source can be
localized to within 10 square degrees as early as a month before
merger at $z = 1$; for higher redshifts, this accuracy is only
possible in the last few days.
\end{abstract}


\maketitle

\section{Introduction}

One of the most promising sources of gravitational waves (GWs) for
LISA is the coalescence of massive black hole (MBH) binaries.
Observations show that massive black holes are present in the center
of nearly all galaxies at the present time \cite{kr95,m98}.
Hierarchical structure formation teaches that these galaxies formed
from the mergers of smaller galaxies and their associated dark matter
halos \cite{m07,dcshs08}.  During such mergers, dynamical friction and
three-body interactions bring the black holes of each galaxy together,
forming a binary.  Eventually the binary is close enough that it
decays due to gravitational radiation \cite{bbr80}.  For $10^4 \,
M_\odot < M < 10^7 \,M_\odot$, these waves lie in the low-frequency
($3 \times 10^{-5} \, \mathrm{Hz} < f < 1 \, \mathrm{Hz}$) band of
LISA.  Merger tree calculations show that $\sim$ tens of events should
be detected by LISA each year \cite{svh07, mhsa07}.

The GWs detected by LISA encode a great deal of information about the
source binary.  In general, GWs from MBH binaries are characterized by
17 parameters.  Since rapid circularization is likely for comparable
mass binaries \cite{p64}, we consider only circular orbits, reducing
the parameter set to 15.  These parameters are the masses of the
holes, their spins, the orientation of the binary orbit, the `merger
time' and `merger phase' (defined by the post-Newtonian
approximation; see \cite{lh06} for details), the position of the
binary on the sky, and its luminosity distance.  LISA estimates of the
sky position and distance, along with the errors in those
measurements, define a three-dimensional `pixel' in which the source
is expected to be located on the sky.  It will be interesting to
search this pixel for electromagnetic (visible, radio, x-ray)
counterparts to the GW source.

It is plausible that no significant electromagnetic activity occurs in
conjunction with MBH coalescence.  In this case, one could imagine
searching the pixel for a galaxy with a structure that indicates a
recent merger.  Another possibility is to search for a galaxy which
has a bulge radial velocity consistent with the GW-measured final
black hole mass.  (This of course assumes that the well-known relation
between these quantities in the local universe \cite{fm00, g00} holds
at high redshift and so soon after a merger.)

It is likely, however, that there is some unique EM activity
associated with the MBH coalescence.  The nature of such activity has
become a hot research topic in recent years, leading to many
possible scenarios \cite{dssch06, khm08}.  For example, if the
surrounding gas is completely swept away by the binary, there may be
no signal during the coalescence.  Instead, there would be a delayed
afterglow when the gas later accretes onto the remnant hole
\cite{mp05}.  This afterglow may occur years after the merger.  It is
likely, though, that the gas will not be totally swept away, leaving enough
to accrete onto the holes and create variable EM activity during the
inspiral phase.  For example, Armitage and Natarajan showed that for a
large mass ratio binary, any gas which does remain will be driven in
during inspiral, producing an EM signal \cite{an02}.  More recent work
by MacFadyen and Milosavljevi\'c showed that periodic variations in the
Newtonian potential can create a quasi-periodic EM flux \cite{mm08}.

Other scenarios predict transients during or immediately after the
coalescence.  One example relies on the fact that the mass of the
black holes is partially radiated away by the gravitational waves.
This causes a near-instantaneous change in the gravitational potential
to which the gas will quickly react, producing shocks \cite{bp07}.
Another possibility is that the `kick', or momentum imparted to the
black hole due to an asymmetric emission of GWs
{\cite{f83,clzm07,sbvbbckm08}} will send the remnant through the
surrounding gas, again producing shocks \cite{lfh08, sb08, sk08}.  A
transient signal might also appear when the GWs are viscously
dissipated in the surrounding gas \cite{kl08}.

Finding a counterpart could greatly enhance the science return of MBH
measurement.  For example, counterparts can improve LISA's ability to
determine certain parameter values.  The sky position is correlated
with various other parameters, particularly luminosity distance and
orbit orientation.  When it is determined exactly by identification of
a counterpart, the other parameters can be estimated to greater
accuracy \cite{h02,hh05}.  Another difficulty with parameter
estimation is that the estimated masses are actually `redshifted';
that is, they are equal to the rest frame mass multiplied by $(1+z)$.
The GWs themselves only give the luminosity distance, not the
redshift, so any decoupling of mass and redshift will be dependent on
a cosmological model \cite{h02}.  The counterpart gives the redshift
directly.

\renewcommand{\thefootnote}{\arabic{footnote}}

Finding the redshift from a counterpart may also allow us to use MBH
events as cosmological distance measures.  Combining the EM-measured
redshift with the GW-measured luminosity distance creates a Hubble
diagram which is calibrated only by general relativity
\cite{s86,hh05}.  In reality, such `standard sirens'\footnote{This
term, coined by Sterl Phinney and Sean Carroll, is appropriate because
GWs can be thought of as analogous to sound waves.} are affected by
weak lensing of the gravitational waves, which impacts measurements of
the luminosity distance \cite{dhcf03,hl05}.  Still, the systematics
affecting MBH GWs should be completely different from those affecting
Type Ia supernova standard candles \cite{p93, rpk95, wgap03} and could
serve as a useful complement to those sources.

Counterparts are also useful for studying the astrophysics of the MBH
coalescence; indeed, the sheer variety of counterpart scenarios shows
how uncertain these processes are.  Specifically, counterparts may
give insight into gas dynamics and accretion.  For instance, GW
measurements of the mass and EM measurements of the luminosity can
be combined to find the Eddington ratio, $L/L_\mathrm{Edd}$
\cite{kfhm06}.  Finally, the counterparts could be used to test
fundamental physics.  If a counterpart features EM variation in phase
with the gravitational wave signal, the two signals can be compared
to test the equivalence of photon and graviton propagation speed.  Any
difference could be explained by a nonzero graviton rest mass
\cite{khm08}.

Given the potential utility of electromagnetic counterparts, it is
important to understand how well LISA will be able to measure sky
position and distance.  This will give astronomers some indication of
LISA's capabilities, as well as provide guidance regarding LISA mission
design decisions.  In particular, it is interesting to see how well
sky position and distance can be measured in advance of merger
\cite{lh08}.  The intention is to analyze LISA data in real time so
that astronomers can point telescopes at the pixel to look for and
then observe the source.  Waiting until after the merger means that
astronomers will miss any precursor or prompt EM events, which might
also make it harder to even identify the correct source.

\section{Preliminaries}

To estimate parameter measurement accuracies, we use the Fisher matrix
formalism \cite{f92, cf94}; details for our analysis are given in
\cite{lh06}.  The Fisher matrix is defined as

\begin{equation}
\Gamma_{ab} = \left(\frac{\partial h}{\partial \theta^a}\left|\frac{\partial h}{\partial \theta^b}\right.\right) \, ,
\end{equation}
where $h$ is the GW signal measured by LISA and $\theta^a$ are the 15
parameters which characterize the signal.  The inner product is
defined by

\begin{equation}
(a|b) = 4\ \mathrm{Re} \left[\int_0^\infty df \frac{\tilde{a}^*(f)\tilde{b}(f)}{S_n(f)} \right] \, .
\end{equation}
Tildes denote Fourier transforms, which we take using the stationary
phase approximation \cite{cf94,pw95}.  $S_n(f)$ is the one-sided noise
power spectral density, which includes instrument noise \cite{lhh00}
and confusion noise from Galactic \cite{nyp01} and extragalactic
\cite{fp03} white dwarf binaries.  The inverse of the Fisher matrix is
the covariance matrix, $\Sigma^{ab}$, the diagonal entries of which
correspond to (squared) parameter errors.  The Fisher matrix estimates
errors in the 'Gaussian approximation', in which the parameter
posterior probability density is represented by a Gaussian around the
most likely values.  This approximation is known to be good for high
signal-to-noise ratio (SNR) \cite{f92, cf94}.  For $z = 5$, our
signals have median SNR $\sim 10$ a month before merger and $\sim
55-370$ at merger, depending on the black hole masses.  For $z = 1$,
this improves to $\sim 60-95$ a month before merger and $\sim
650-2700$ at merger.  However, it is not immediately clear if these
values are large enough for a signal with 15 strongly coupled
parameters.

We use only the inspiral portion of the coalescence waveform.  This is
largely because the inspiral can be modeled by the post-Newtonian
approximation to general relativity \cite{b06,bdiww95,ww96}, offering
a convenient parameterization for our analysis.  In addition, the
primary source of information about sky position is the motion of LISA
around the Sun.  Therefore, one might guess that the inspiral provides
the bulk of the localization.  However, with the development of
numerical relativity, it has become possible to perform similar
calculations which include the merger phase of the waveform.
Preliminary results seem to show that the merger could dramatically
improve sky position and distance determination \cite{bhhs08,tmkfab08}.

Our waveforms are carried out to second post-Newtonian (2PN) order in
the phase, but include only the Newtonian quadrupole amplitude term.
They also include the effects of spin-induced precession.  Spins
precess due to post-Newtonian spin-orbit and spin-spin coupling
effects.  For example, to the order we are considering
\cite{acst94,k95},

\begin{eqnarray}
\fl
\qquad\quad
\mathbf{\dot{S}}_1 =
\frac{1}{r^3}\left[\left(2+\frac{3}{2}\frac{m_2}{m_1}\right)\mu
\sqrt{Mr}\mathbf{\hat{L}}\right] \times \mathbf{S}_1
+ \frac{1}{r^3}\left[\frac{1}{2}\mathbf{S}_2-\frac{3}{2}(\mathbf{S}_2\cdot
\mathbf{\hat{L}})\mathbf{\hat{L}}\right] \times \mathbf{S}_1 \, ,
\end{eqnarray}
where $m_1$ and $m_2$ are the black hole masses, $M = m_1 + m_2$ is
the total mass, $\mu = m_1m_2/M$ is the reduced mass, $\mathbf{S}_1$
and $\mathbf{S}_2$ are the spin vectors, $\mathbf{\hat{L}}$ is the
normal to the orbital plane, and $r$ is the orbital separation in
harmonic coordinates.  On short timescales, total angular momentum is
conserved, so the orbital plane precesses to compensate for the spins.
The effect of these precessions on the waveform is to create various
amplitude and phase modulations.  For example, consider the
`polarization amplitude', originally introduced by Cutler
\cite{c98}. It is constructed by adding the two polarizations,
weighted by their antenna pattern functions $F^+$ and $F^\times$, in
quadrature.  Various parts of the resulting GW amplitude (varying by
author, but always including the geometrical factors) are then defined
to be the polarization amplitude.  We use the convention of Vecchio
\cite{v04}:

\begin{equation}
A_{\mathrm{pol},i} = \frac{\sqrt{3}}{2}[(1+(\mathbf{\hat{L}}\cdot \mathbf{\hat{n}})^2)^2 F_i^+(t)^2 + 4(\mathbf{\hat{L}}\cdot \mathbf{\hat{n}})^2 F_i^\times(t)^2]^{1/2} \, , 
\end{equation}
where $\mathbf{\hat{n}}$ is the direction to the binary and $i \in
\{\mathrm{I,II}\}$ denotes one of two detectors synthesized from the
LISA laser links.  Because $\mathbf{\hat{L}}$ precesses, an amplitude
modulation is imposed on the measured waveform.  Other modulations are
hidden inside the antenna pattern functions.

\begin{figure}[ht]
\begin{center}
\includegraphics[scale=0.5]{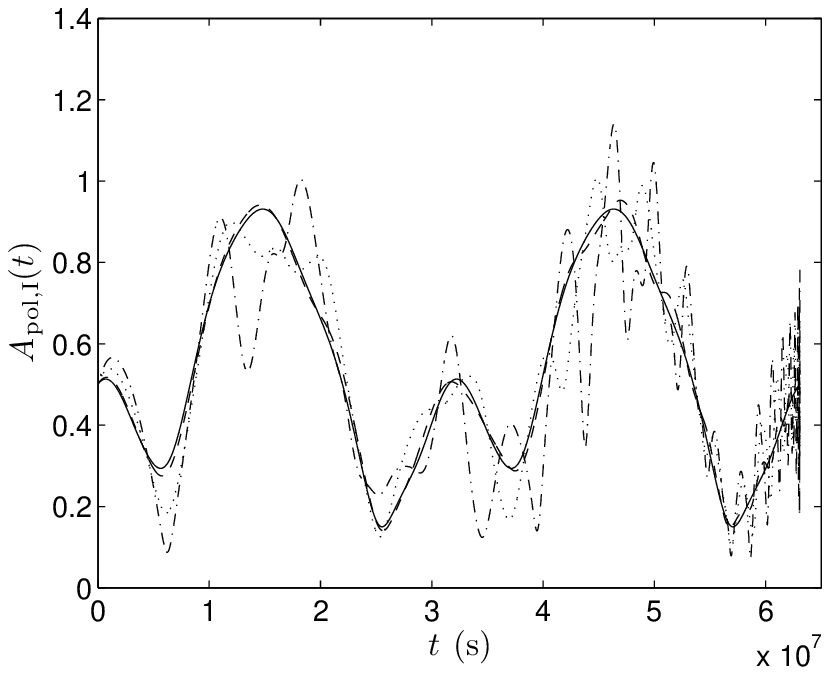}
\caption{`Polarization amplitude' $A_{\mathrm{pol}}(t)$ of the
signal measured in detector I for a binary with $m_1 = 10^6 M_\odot$,
$m_2 = 3\times 10^5 M_\odot$, and $z = 1$.  The curves are as follows:
solid line, $\chi_1 = \chi_2 = 0$; dashed line $\chi_1 = \chi_2 =
0.1$; dotted line, $\chi_1 = \chi_2 = 0.5$; and dash-dotted line,
$\chi_1 = \chi_2 = 0.9$.  ($\chi = S/m^2$ is the dimensionless spin
parameter.) The figure covers the final two years of inspiral.
Without spin, the only modulation is due to LISA's orbit around the
Sun, which is the primary source of localization information.
Precession adds additional modulations, which increase toward merger.
(This figure appeared previously as figure 1 of \cite{lh06}; however,
an error in the code \cite{lh06e} caused it to be incorrect in the
published version.)}
\label{fig:polamp}
\end{center}
\end{figure}

Figure \ref{fig:polamp} shows $A_{\mathrm{pol},i}$ for a two-year
cycle.  Without precession, the polarization amplitude has a period of
one year, the time it takes LISA to orbit the Sun.  This orbital
modulation (which is hidden in the pattern functions $F^+$ and
$F^\times$) plays a large role in resolving the position of a source
on the sky.  When precession is added, the amplitude is modulated much
more strongly.  The additional modulation is strongest just before
merger, when the holes are closer and the post-Newtonian precession
effects the largest.  Similar modulations exist in the relative phase
between weighted polarizations (the `polarization phase') and the
intrinsic post-Newtonian phase.  Finally, there is an additional
precessional correction to the orbital phase \cite{acst94}.  All of
these effects combine to provide more information about the source,
reducing parameter errors.  Specifically, they break degeneracies
between strongly correlated parameters, notably the sky position,
distance and orbital orientation.  We expect the improvements to be
greatest when the precession effects are largest, at the end of
inspiral.

\section{Time evolution of LISA pixel}
The operation of our code is described in \cite{lh06}.  We first
select rest frame masses and redshift for a binary system.  We then
randomly choose the spin magnitudes and direction, orientation, sky
position and post-Newtonian `merger time' parameter $t_c$.  For
each choice, the code computes the Fisher matrix $\Gamma_{ab}$ and the
covariance matrix $\Sigma^{ab}$.  Extracting the sky position error
ellipse from this requires some care.  Defining the ellipse such that
the probability of the source lying outside it is $e^{-1}$ ($\sim
63\%$ confidence interval) \cite{c98}, the major ($2a$) and minor
($2b$) axes of that ellipse are given by

\begin{eqnarray}
\fl
\left\{
\begin{array}{c}
2a \\
2b
\end{array}
\right\} = 2\left[
\vphantom{\sqrt{(\Sigma^{\bar{\mu}_N\bar{\mu}_N} -
\Sigma^{\bar{\phi}_N\bar{\phi}_N})^2 +
4(\Sigma^{\bar{\mu}_N\bar{\phi}_N})^2}} \csc^2 \bar{\theta}_N
\Sigma^{\bar{\mu}_N\bar{\mu}_N} + \sin^2 \bar{\theta}_N
\Sigma^{\bar{\phi}_N\bar{\phi}_N} \right. \nonumber \\ \left. \pm
\sqrt{(\csc^2 \bar{\theta}_N \Sigma^{\bar{\mu}_N\bar{\mu}_N} - \sin^2
\bar{\theta}_N \Sigma^{\bar{\phi}_N\bar{\phi}_N})^2 +
4(\Sigma^{\bar{\mu}_N\bar{\phi}_N})^2} \right]^{1/2} \, ;
\label{eq:2a2b}
\end{eqnarray}
the upper sign is for $2a$, while the lower sign is for $2b$.  Here
$\bar{\mu}_N = \cos \bar{\theta}_N$, where $\bar{\theta}_N$ is the
polar angle of the source.  $\bar{\phi}_N$ is the azimuthal angle.
The bars over the angles reflect that they are measured in the
barycenter frame, rather than in a frame moving with LISA.  (See
\cite{lh06} for more details.)

We find that at merger, the sky position error ellipse has a (median)
major axis of $\sim 15-45$ arcminutes at $z = 1$, depending on system
mass.  This figure increases to $\sim 3-5$ degrees at $z = 5$.  The
minor axis ranges from $\sim 5-20$ arcminutes at $z = 1$ to $\sim 1-3$
degrees at $z = 5$.  Finally, the luminosity distance errors $\Delta
D_L/D_L$ are $\sim 0.002-0.007$ at $z = 1$ and $\sim 0.025-0.05$ at $z
= 5$.  It is important to note that these luminosity distance errors
do not include the effects of weak lensing.  Lensing due to
intervening matter will magnify or demagnify the waves, causing an
incorrect measurement of luminosity distance.  The distance error 
scales with redshift roughly as $\Delta D_L/D_L \simeq 0.044 z$ for low $z$
\cite{hl05}.  It is expected that this dependence will become flat
at some transition redshift, most likely near $z \sim 3$ or $4$; the
precise transition depends upon the (poorly understood) high redshift
mass function.  With the development of high-quality weak lensing
maps, one might think that it would be possible to correct for the
impact of lensing and recover much of the intrinsic GW distance
measurement precision.  Unfortunately, lensing noise arises mostly from
structure on subarcminute scales that is not probed by shear maps, making any substantial correction impossible \cite{dhcf03}.

A complete listing of the time-dependent results can be found in 
\cite{lh08}.  To help summarize these results, we take an area of 
10 square degrees as a useful benchmark for a
`good' localization in the plane of the sky.  This is roughly the 
field of view of the Large Synoptic Survey Telescope (LSST) \cite{t02}.  
Then for low redshift ($z = 1$), sources are well localized as far
back as a month before merger for most masses.  At $z = 3$, this 
condition is met a few days before merger, but only for a subset of 
the masses we consider.  Finally, at $z = 5$, the error 
ellipse is the size of the LSST field or smaller at most a day before 
merger, and only then in very few cases.

At $z = 1$, the luminosity distance errors increase to $\sim 2-4 \%$ a month before merger in most cases.  Lensing should dominate the error budget for all but the highest masses, for which the source is in band too short a time to be well localized (if it is in band yet at all).  At $z = 3$, the distance errors are as large as $\sim 10\%$ or more a month before merger; $\sim 20 \%$ becomes more typical at $z = 5$.  Since the lensing error flattens out with redshift, these intrinsic errors might dominate, but the high-$z$ lensing noise is too uncertain to make a definite statement.

\begin{figure}[h]
\begin{center}
\includegraphics[scale=0.5]{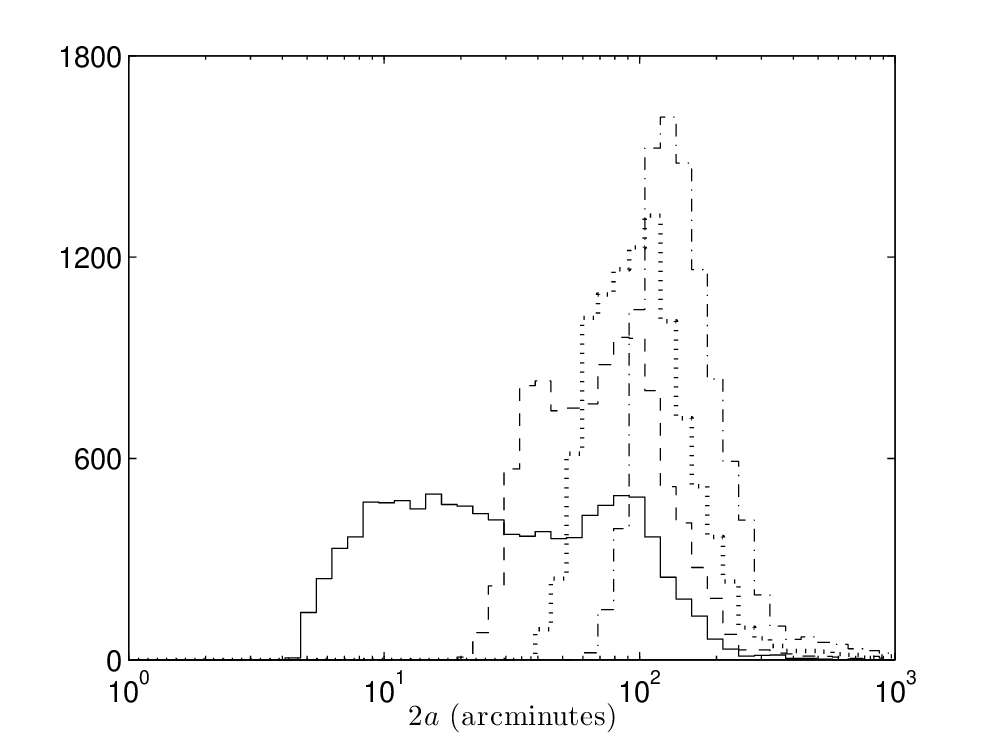}
\includegraphics[scale=0.5]{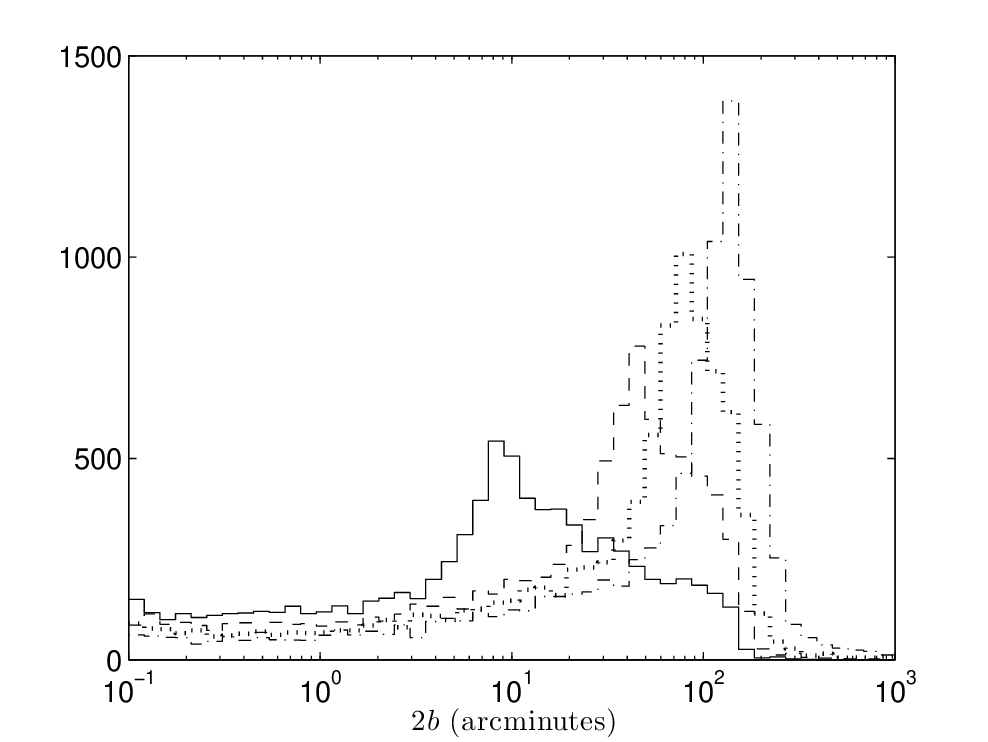}
\caption{Distribution of the major axis $2a$ (top) and minor axis $2b$
(bottom) of the sky position error ellipse for $10^4$ binaries with
$m_1 = 10^6 M_\odot$, $m_2 = 10^5 M_\odot$, and $z = 1$ at different
times before merger.  Reading from left to right, the number of days
until merger is 0 (solid line), 1 (dashed line), 7 (dotted line), and
28 (dash-dotted line).  The largest change occurs in the final day
before merger.}
\label{fig:2a2bevol}
\end{center}
\end{figure}

Figure \ref{fig:2a2bevol} shows the evolution of the major axis and
minor axis of the sky position error ellipse for $m_1 = 10^6 M_\odot$,
$m_2 = 10^5 M_\odot$ and $z = 1$.  Each histogram represents $10^4$
different choices of the random parameters.  As the time until final
merger decreases from 1 month to 1 day, the median error gradually
decreases.  In addition, the shape of the distribution changes
slightly, becoming less peaked.  However, a drastic change occurs
between 1 day and 0 days (end of inspiral).  In just this day, the
median is reduced by half an order of magnitude and the shape of the
distribution changes drastically.  For the major axis, the
distribution becomes almost flat over a certain range and displays a
slightly bimodal structure.

\begin{figure}[ht]
\begin{center}
\includegraphics[scale=0.5]{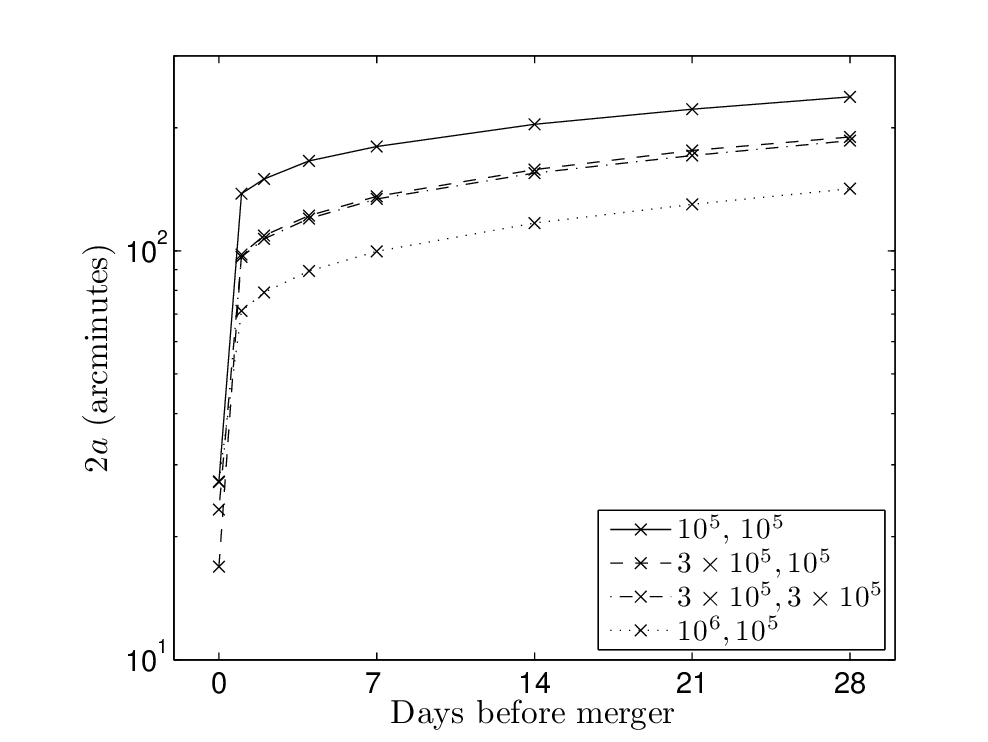}
\includegraphics[scale=0.5]{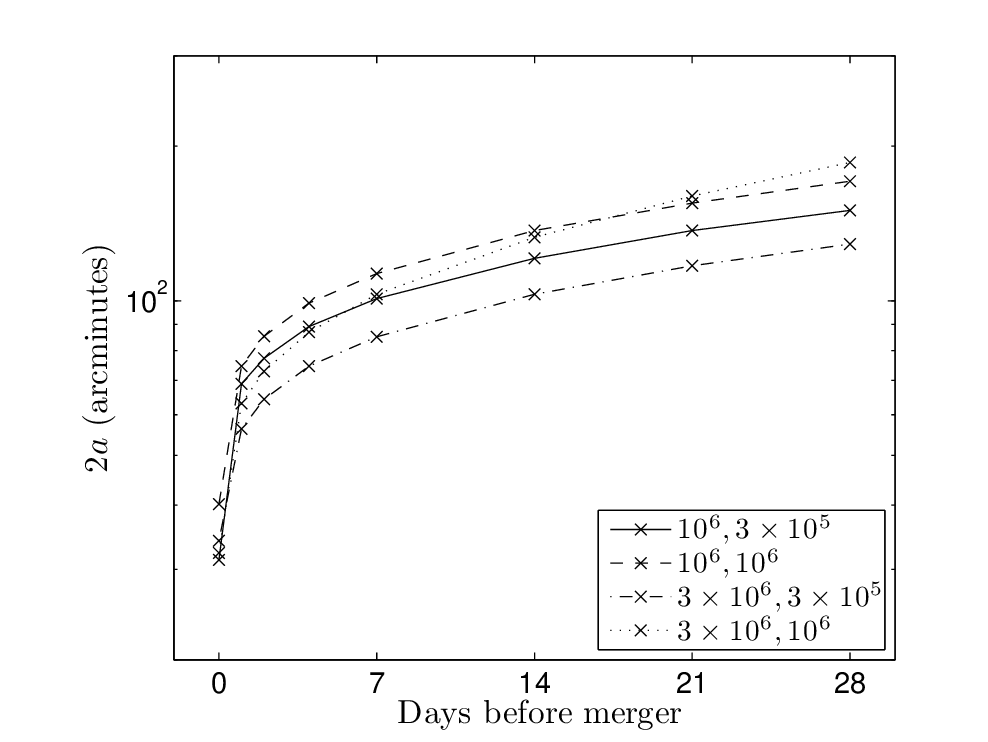}
\caption{Median values of $2a$, the sky position ellipse major axis,
as a function of time before merger.  Each point is taken from a Monte
Carlo sample of $10^4$ binaries with fixed masses (given in units of
$M_\odot$) and $z = 1$.  Data were only output at the marked points;
the lines are there to guide the eye.}
\label{fig:medians}
\end{center}
\end{figure}

\begin{figure}
\begin{center}
\includegraphics[scale=0.5]{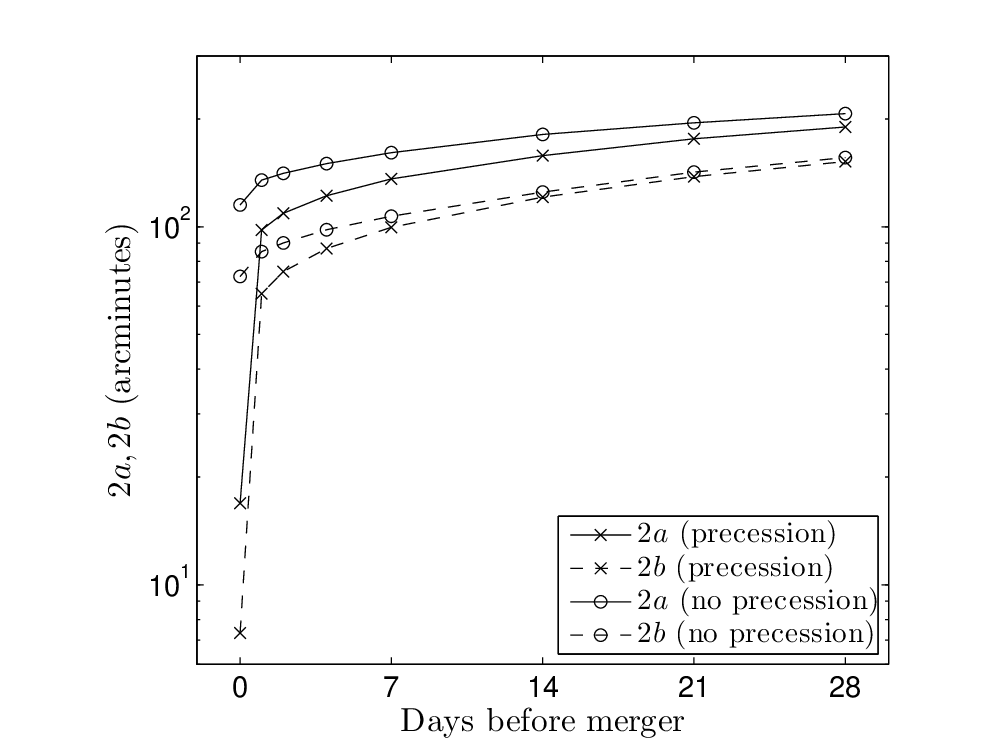}
\caption{Medians of $2a$ (solid lines) and $2b$ (dashed lines) as a
function of time for binaries with $m_1 = 3 \times 10^5 M_\odot$, $m_2
= 10^5 M_\odot$ and $z = 1$.  Spin-induced precession was included in
the waveform for the crosses and neglected for the circles.}
\label{fig:NPcomparison}
\end{center}
\end{figure}

Figure \ref{fig:medians} shows the median values of $2a$
for different masses.  We see a trend holding over all masses: gradual
improvement with time until a sharp change the last day.  The change
is greatest for the lowest mass systems.
These results agree quite well with those of Kocsis \etal
\cite{khmf07}, except their study does not show a sharp improvement in
localization during the final day of inspiral.  It turns out that this
effect is due mostly to spin precession, which Kocsis \etal do not
include.  This late impact of precession is to be expected following
our discussion of figure \ref{fig:polamp} and is confirmed in figure
\ref{fig:NPcomparison}.  By breaking correlations between parameters,
precession gives a factor of $2-7$ improvement in final parameter
estimation errors.  However, since most of this improvement takes
place in the final days before merger, precession does not help very
much with advanced localization.  This effect may, however, impact the
frequency of data transfer between LISA and the ground: as a source
approaches merger, it may be useful to continuously downlink data to
take advantage of the rapid improvement in localization.

As a final study, we also investigate the position dependence of the
localization pixel.  Rather than randomly choosing the full sky
position, we randomly choose just one of the polar ($\bar{\theta}_N$)
and azimuthal ($\bar{\phi}_N$) angles, while varying the other in a
controlled fashion.  It turns out that only the polar dependence is
meaningful due to a particular feature of our formalism: we randomly
choose the final merger time within LISA's mission window, meaning
that even with a specifically chosen $\bar{\phi}_N$, the relative
azimuth between LISA and the source at merger, $\delta \phi =
\bar{\phi}_N - \phi_{\mathrm{LISA}}(t_c)$, is random.  This mostly
washes out the azimuthal dependence.  Other authors
\cite{khmf07,ts08,khm08} make a different choice, in which the
position of LISA at merger is fixed.  This allows them to study the
position dependence of localization at a particular time of the year;
consequently, their results for both polar and azimuthal angle
dependence are different from ours.

We find that the sky position axes and area are maximal for sources in
the ecliptic plane and minimal near the poles, at $\cos \bar{\theta}_N
= \pm 0.8$.  The distance errors have a three-peaked structure, with
one peak at the ecliptic plane and the others very close to where the
sky position errors are minimized.  Converting to Galactic
coordinates, we find that some of the best localized sources lie
outside the Galactic plane.  This suggests that the most easily
localized binaries will not tend to lie in regions of the sky with
thick foregrounds or high levels of extinction.

\section{Summary and future work}

Observing electromagnetic counterparts to massive black hole binary
coalescences can be very useful in the contexts of parameter
estimation, cosmology, astrophysics and fundamental physics.  To
observe a counterpart, and to get the most science out of it, it is
useful to be able to localize the source prior to merger.  Our study
shows that advanced localization of sources with LISA should be
possible at low redshift ($z = 1$), but begins to be difficult at
higher $z$.  The effects of precession break degeneracies between
parameters, dramatically improving estimates in the final days before
merger.  Finally, at least some of the best located sources will be
out of the Galactic plane, improving our chances of identifying
counterparts.

These results are not the final story on LISA's ability to measure and
localize MBH sources.  For example, it is useful to test whether the
Gaussian approximation/Fisher matrix approach is actually valid for a
signal with 15 parameters.  This problem is currently being approached
both by analytic analysis \cite{v08} and by comparing the Fisher
results to errors generated by a Markov Chain Monte Carlo exploration
of the parameter posterior distribution function \cite{cc05}.  Another
approximation worth checking is the stationary phase method for
obtaining Fourier transforms; this is known to be good for signals
without precession \cite{dkpo99} but seems to smooth out sharp
features caused by strong precession \cite{colh}.  Finally, there are
various extra pieces which can still be added to the waveform,
including higher post-Newtonian phase terms, higher order precession
equations \cite{fbb06, bbf06}, higher post-Newtonian amplitude terms,
and harmonics beyond the quadrupole.  The latter two terms, often
called simply `higher harmonics', have already been shown to
decrease parameter errors in the absence of spin precession
\cite{aissv07, ts08, pc08}.  The improvement seems to be somewhat
orthogonal to the improvement caused by precession, but a more
complete investigation, including time dependence, is necessary.  We
and our collaborators are currently investigating these issues.

\section*{Acknowledgments}

Our work on LISA measurement analysis has been supported by NASA
Grants NNG05G105G and NNX08AL42G, as well as NASA contract no.\
1291617 and the MIT Class of 1956 Career Development Fund.  We thank
Daniel Holz for attempting to educate us on the redshift dependence of
weak lensing distance errors.  We also gratefully thank the organizers of
the Symposium for the travel support provided to RNL and for the
outstanding meeting that they put together.

\end{document}